\documentclass[12pt]{article}
\usepackage{color,amsmath,amssymb,amsthm,bm,epsf,epsfig,enumerate,alltt,latexsym,graphicx}
\usepackage{comment}
\newtheorem{propn}{Proposition}

\newcommand{\pr}{\ensuremath{^{\prime}}}
\newcommand{\id}{\ensuremath{\mathrm{I}}}
\renewcommand{\P}{\ensuremath{\mathbf{P}\!}}
\renewcommand{\sp}{\ensuremath{\mathrm{sp}}}
\newcommand{\tr}{\ensuremath{\mathrm{tr}}}

\newcommand{\Diag}{\ensuremath{\mathrm{Diag}}}
\newcommand{\mbk}{\ensuremath{\mathbb{K}}}

\newcommand{\bc}[1]{\ensuremath{\mathbf{\mathcal{#1}}}}

\begin{document}
\begin{center}
\begin{Large}
{\bfseries A Formula for Type III Sums of Squares}\end{Large}\\
\vspace{.5cm}by Lynn R. LaMotte\footnote{School of Public Health, LSU Health, New Orleans, LA USA 70124,  {\tt llamot@lsuhsc.edu}}
\end{center}

\begin{abstract}
Type III methods were introduced by SAS to address difficulties in dummy-variable models for effects of multiple factors and covariates.  They are widely used in practice; they are  the default method in several statistical computing packages.  Type III sums of squares (SSs) are defined by a set of instructions; an explicit mathematical formulation does not seem to exist.  

An explicit formulation is derived in this paper. It is used to illustrate Type III SSs and their properties in the two-factor ANOVA model.
\end{abstract}

\vspace{.5cm}\noindent{\sc Key Words}: Type III Effects, Unbalanced Two-Factor ANOVA, Yates's MWSM

\section{Introduction}
Type III estimable functions and sums of squares (SSs) came to light in SAS publications in the 1970s, mainly Goodnight (1976) and SAS (1978).   They are defined by  instructions to construct  a set of contrasts for a named effect.  These become the Type III estimable functions, and the Type III SS is the squared norm of the orthogonal projection of the response vector onto the linear subspace that they define.  That becomes the numerator SS for an $F$-statistic for testing the effect.

At that time, there was considerable discussion about how to define SSs to test main effects when treatments are defined by combinations of levels of multiple factors.  Such questions grew out of R. A. Fisher's Analysis of Variance (ANOVA)  (see Fisher 1938).  
F. Yates (1934) introduced the \emph{method of weighted squares of means } (MWSM) to provide  SSs for testing main effects of two factors in unbalanced models that included possible interaction effects.  He did not claim any properties for it, other than that it ``provides an efficient estimate ... of the variance of the individual observations.''  He gave no rationale for its definition, and he did not attempt to describe how it might be extended to more general settings.  In the ensuing decades, it became a touchstone for methods to define SSs for factor main effects.  It provided an explicit formula, so that the SS could be computed on the desktop calculators available in 1934.  However, its properties are not readily apparent from the formula itself. 

There was not consensus on the definitions of effects in unbalanced models. Kutner (1974) lists three definitions of main effects, and Speed et al. (1978) list four ``common ANOVA hypotheses'' that define main effects.  SAS's motivation seems to have been to provide an omnibus approach that reduced to best practice where it was known to exist.  Goodnight (1976) asserted, ``When no missing cells exist in a factorial model, Type III SS will coincide with Yate's weighted squares of means technique.'' That assertion is repeated in SAS documentation, including the latest version, and phrasing like it appears in other sources.

Type III SSs were implemented by other statistical computing packages.  As they became widely known and available, then, as far as practitioners were concerned, the questions about what SS to use in unbalanced models or models with covariates had been addressed and resolved, somehow. Conventional wisdom became, ``use Type III.''

It appears that one could follow the instructions given in SAS documents and documentation to program Type III SSs in a general setting. Doing so is not altogether simple, though.  The algorithm available in the {\tt car} package in R (D. Wollschl\"ager {\tt www.dwoll.de/r/ssTypes.php}, 2017), for example, does not always give the same results as SAS does.

As far as I have been able to find, no explicit, concise mathematical formulation of Type III SSs has appeared.  Some descriptions in print give the impression that they can be had as extra SSE (Error SS) due to deleting sets of columns when ANOVA terms are represented by contrasts instead of dummy variables (Venables 2000, p. 15, Mangiafico 2015, for example). While that is correct in some cases, it is not in general, and so it does not define Type III SSs.  Others (Searle 1987, p. 463, and Langsrud 2003) have asserted that Type III SSs result from imposing the ``usual'' zero-sum restrictions on regression coefficients, but that is not correct in general either.

While there are many descriptions that assert features of Type III SSs, they do not define them. In addition to SAS's step-by-step instructions, though, the description given in SPSS documentation is a succinct definition.  From SPSS Statistics $>$ SPSS Statistics 20.0.0 $>$ Help $>$ Statistics Base Option $>$ GLM Univariate Analysis $>$  GLM Model, it is:
\begin{quotation}
\noindent This method calculates the sums of squares of an effect in the design as the sums of squares adjusted for any other effects that do not contain it and orthogonal to any effects (if any) that contain it.
\end{quotation}
In Section \ref{Type III} a concise mathematical formulation of Type III SS is derived from this definition.

Properties of Type III SSs are difficult to establish without an explicit mathematical formulation.  As far as I have been able to find, no proofs of any properties have appeared, and beliefs about them are based on observation and experience. It seems to be widely believed that Type III SSs test classical ANOVA effects when those effects are estimable. The source of that belief appears to be the assertion by Goodnight (1976) quoted above.  Khuri (2010, Section 10.3) says ``[i]t can be shown'' that the assertion is true, citing ``Speed and Hocking, 1976, p. 32; Searle, Speed, and Henderson, 1981, Section 5.2; Searle, 1994, Section 3.1.''  No proof appears in any of these sources, or in sources that they cite.  Searle (1994, Section 3.1), for example, says, ``It is well known for cross-classified fixed-effects models with all-cells-filled data that the SAS Type III sums of squares are those of Yates's (1934) weighted squares of means analysis.'' He gave no proof and cited no source. Speed and Hocking (1976, after (11)) say that ``closed-form solutions are not readily available'' and that ``the interested reader is encouraged to verify these points'' with ``a set of data with unequal numbers of observations per cell.''

The formulation of Type III SSs is applied in Section \ref{Illustration} to the unbalanced two-factor ANOVA model. An explicit expression for the SS for A main effects is derived for the all-cells-filled case, and it is used to prove the properties mentioned above. 

The Type III approach has been criticized in strong words.  Milliken and Johnson (1984, p. 185) say, when there are empty cells, ``... we think that the Type III hypotheses are the worst hypotheses to consider ... because there seems to be no reasonable way to interpret them.''  Venables (2000, p. 12) says, ``I was profoundly disappointed when I saw that S-PLUS 4.5 now provides `Type III' sums of squares as a routine option ... .''  The debate on the merits of Type III methodology continues: see Macnaughton (1998), Langsrud (2003), Hector et al. (2010), and Smith and Cribbie (2014).  

Notation and definitions:  For matrices $A$, $B$, $C$, and $D$ with dimensions so that the operations are defined, matrix transpose, sum, and product are denoted $A\pr$, $A+B$, and $AB$. 
The Kronecker product of $A$ and $B$, denoted $A\otimes B$, is the matrix formed by replacing each entry $a_{ij}$ of $A$ by the matrix $a_{ij}B$.  There are many useful properties and relations for Kronecker products; a few that are used implicitly here are: $(A\otimes B)\pr = A\pr \otimes B\pr$,  $A\otimes (B+C) = A\otimes B + A\otimes C$, $(A+B)\otimes C = A\otimes C + B\otimes C$, $(A\otimes B)(C\otimes D) = (AC)\otimes (BD)$.  The sum of the diagonal entries of a square matrix $A$ is the \emph{trace} of $A$ and denoted $\tr(A)$.  If the operations are defined, $\tr(A) = \tr(A\pr)$, $\tr(AB) = \tr(BA)$, and $\tr(A\otimes B) = \tr(A)\tr(B)$.  The matrix formed by concatenating the columns of $A$ and $B$ (both with $r$ rows) is denoted $(A, B)$.
 
The linear subspace spanned by the columns of $A$ (the column space of $A$) is denoted by $\sp(A)$. Its orthogonal complement is denoted $\sp(A)^\perp$.  The orthogonal projection matrix onto $\sp(A)$ is denoted $\P_A$. It is symmetric and idempotent, and it is the unique matrix such that, for each $r$-vector $\bm{y}$, $\P_A\bm{y}$ is the orthogonal projection of $\bm{y}$ in $\sp(A)$.  The uniqueness can be stated as: $\P_A = \P_B$ iff $\sp(A) = \sp(B)$.
$\P_A$ can be computed as $Q_AQ_A\pr$, where $Q_A\pr Q_A = \id_{\nu}$ and $\sp(Q_A) = \sp(A)$. A representation used frequently in the derivations here is that $\P_A = A(A\pr A)^-A\pr$, where $(A\pr A)^-$ is a generalized inverse of $A\pr A$. A property useful here is that $\P_{A\otimes B} = \P_A\otimes \P_B$.

\section{A Formula for Type III SS}\label{Type III}

It is assumed here that $\bm{Y}$ (a column vector) follows an $n$-variate normal distribution with mean vector $\bm{\mu} = X\bm{\beta}$ and variance-covariance matrix $\sigma^2\id$. Its realized value is $\bm{y}$. The $n\times k$ model matrix $X$ is fixed and known.   The unknown parameters of its distribution are $\bm{\beta}$, a $k$-vector, and $\sigma^2$, a positive scalar. The model for the mean vector is the set of possibilities for $\bm{\mu}$. It is $\{X\bm{\beta}: \bm{\beta}\in\Re^k\}$, the set of all linear combinations of the columns of $X$. Equivalently, it is $\sp(X)$.  Conventional usage is to refer to the model simply as $X\bm{\beta}$.

In the general ANOVA framework, the $n$ subjects are observed under combinations of levels of multiple factors and values of covariates. A linear model for the population means can be formulated in terms of indicator variables -- usually called \emph{dummy} variables -- for the factor-level combinations. A model can be specified by a list of names of factor effects (main effects, interaction effects), covariates, and factor-by-covariate effects.    It takes the general form $X\bm{\beta}$, with sets of columns of $X$ identified with the list of effect names.   There is a partial ordering of the names of effects called \emph{containment}. Type III for the name of a given target effect  in the model is based on the partition of columns of $X$ corresponding to names of effects that do not contain the target, those that do contain it, and the target itself.

Corresponding to the description quoted above from SPSS, consider the columns of $X$ to be partitioned as $X=(X_0, X_1, X_2)$, and consider $\bm{\beta}$ to be partitioned accordingly as $(\bm{\beta}_0\pr, \bm{\beta}_1\pr, \bm{\beta}_2\pr)\pr$.  $X_1$ is defined by the ``effect in the design'' that is the target of interest.    $X_0$ is defined by names of ``effects that do not contain it,'' and $X_2$ by names of  ``any effects (if any) that contain it.''  

While  ``contain'' as used in this definition is well-defined,  it plays no role in this section.  However, in dummy-variable formulations of $X$, terms for any given effect generate a linear subspace that contains (in the point-set sense) the linear subspaces generated by any other effects that it contains (in the sense meant in the definition).

The word ``effects'' in the definition refers  to sets of estimable functions.  The estimable linear functions of $\bm{\beta}$ take the form $\bm{h}\pr X\bm{\beta}$, where $\bm{h}$ is an $n$-vector.   
Assume that $\bm{h}\in\sp(X)$, that is, that $\bm{h}=\P_X\bm{h}$.  That $\bm{h}$ be ``adjusted for $X_0$'' requires that $X_0\pr\bm{h}=\bm{0}$, or $\bm{h}\in\sp(X_0)^\perp\cap\sp(X)$.

Estimable functions of $\bm{\beta}_2$ are those that do not involve $\bm{\beta}_0$ or $\bm{\beta}_1$, that is, $\bm{m}\pr X\bm{\beta}$ with $X_0\pr\bm{m} = \bm{0}$ and $X_1\pr\bm{m}=\bm{0}$.  Equivalently, $\bm{m}\in\sp(X_0, X_1)^\perp\cap\sp(X)$. Let $N_{01}$ be a matrix such that $\sp(N_{01})=\sp(X_0, X_1)^\perp\cap\sp(X)$.  Then estimable functions of $\bm{\beta}_2$ are $\bm{m}\pr X\bm{\beta}$ with 
\[X\pr\bm{m} = \left(\begin{array}{c}0\\ 0\\ X_2\pr\bm{m}\end{array}\right) = 
\left(\begin{array}{c}0\\ 0\\ X_2\pr N_{01}\bm{c}\end{array}\right)
\]
for some $\bm{c}$. That an estimable function $\bm{h}\pr X\bm{\beta}$ be orthogonal to all estimable functions $\bm{m}\pr X\bm{\beta}$ of $\bm{\beta}_2$ requires that
\[ (X\pr\bm{h})\pr (X\pr\bm{m}) = \bm{h}\pr X_2(X_2\pr N_{01}\bm{c}) = 0
\]
for all vectors $\bm{c}$, which requires that $\bm{h}\in \sp(X_2X_2\pr N_{01})^\perp$.

Let $X_{2*} = X_2X_2\pr N_{01}$.  Putting these together, the Type III estimable functions  are  $\{(X\pr\bm{h})\pr \bm{\beta}: \bm{h}\in \bc{S}_{3}\}$,
where 
\begin{equation}\label{S3}
\bc{S}_{3}=\sp(X_0, X_{2*})^\perp\cap\sp(X).
\end{equation}

Let $\P_{3}$ be the orthogonal projection matrix onto the linear subspace $\bc{S}_{3}$, so that $\P_3 = \P_X - \P_{(X_0, X_{2*})}$. 
 Given an $n$-vector of realized values $\bm{y}$, the Type III numerator SS is $SS_3=\bm{y}\pr \P_{3}\bm{y}$, and its df (degrees of freedom) is $\tr(\P_{3})$. Its ncp (non-centrality parameter) is $\bm{\delta}_3\pr\bm{\delta}_3/\sigma^2$, where $\bm{\delta}_3 = \P_3X\bm{\beta}$, and it is 0 iff $\P_3X\bm{\beta}=\bm{0}$.

We shall say that a (numerator) SS \emph{tests exactly} H$_0: G\pr\bm{\beta}=\bm{0}$ (or simply $G\pr\bm{\beta}$) iff its ncp is 0 iff $G\pr\bm{\beta}=\bm{0}$. In this sense, $SS_3$ tests exactly $\P_3X\bm{\beta}$.

$\P_3$ can be computed in several ways. One way is in two steps, with the Gram-Schmidt (GS) construction as described in LaMotte (2014). From GS on $(X_0, X_1, X)$, take $N_{01}$ as the columns in the orthonormal spanning set contributed by $X$ after $(X_0, X_1)$. Compute $X_{2*}$, then compute $Q_3$ as the columns in the orthonormal spanning set from  GS on $(X_0, X_{2*}, X)$ contributed by $X$ after $(X_0, X_{2*})$. Then $\P_3 = Q_3Q_3\pr$, and its df is the number of columns in $Q_3$.

$\bc{S}_3$ defines the set of all Type III estimable functions generated by the target name and its containment relations to the rest of the model.   The direct role of the target name (to which $X_1\bm{\beta}_1$ corresponds)  seems to be peripheral, appearing only through $N_{01}$ in $X_{2*}$. The construction is driven mainly by the rest of the names in the model, those that contain and those that do not contain  the target name. Like Michelangelo, it trims away everything else to reveal the object of interest.

Conventionally, we would define the effect of interest as a set of linear functions of $\bm{\beta}$, say $G\pr\bm{\beta}$.  The null hypothesis H$_0$ would then be that $\bm{\beta}$ is such that all of these functions are zero. Then we would derive a numerator SS as the restricted model - full model (RMFM) difference in Error SS (SSE), with $X\bm{\beta}$ as the full model and $\{X\bm{\beta}: G\pr\bm{\beta}=\bm{0}\}$ as the restricted model.  

Here, the construction of the test statistic is driven entirely by the definition of Type III estimable functions. The effect in question (the conditions imposed to get the restricted model) is not defined directly.   The construction defines a sum of squares. It is not generated by any hypothesis.

However, note that $SS_3$ is an RMFM SS. The full model is $\sp(X)$ and the restricted model is $\sp(X_0, X_{2*})$. It can be shown that $\sp(X)$ is the direct sum of $\sp(X_0, X_{2*})$ and $\sp(X_{1|0})$, with $X_{1|0} = (\id - \P_{X_0})X_1$. Then $\sp(X_0, X_{2*})$ is the restricted model resulting from imposing the conditions $X_{1|0}\bm{\beta}_1=\bm{0}$ on the full model. In other words, $SS_3$ tests H$_0: X_{1|0}\bm{\beta}_1 = \bm{0}$. (This is the same hypothesis that so-called Type II SS tests, but Type II has $\sp(X_0, X_1)$ as the full model and $\sp(X_0)$ as the restricted model.  It may be seen from this that Type III df and Type II df are the same.)

Now we can see a rationale for transforming $X$ to $X_* = (X_0, X_1, X_{2*})$.  If $\sp(X)$ is not the direct sum of $\sp(X_0, X_1)$ and $\sp(X_2)$ (if, e.g., $\sp(X_2)\supset\sp(X_1)$), then $X_{1|0}\bm{\beta}_1$ might not be testable (estimable) in the full model.  It is testable in $\sp(X_0, X_1)$.  Without changing $X_0$ or $X_1$, if we can re-express $\sp(X)$ as a direct sum of $\sp(X_0, X_1)$ and another linear subspace, say $\bc{S}_{2\backslash 01} = \sp(X_{2\backslash 01})$, then $X_{1|0}\bm{\beta}_1$ will be testable in the re-expressed full model. 

The choice of $\bc{S}_{2\backslash 01}$ (and the matrix $X_{2\backslash 01}$ used to generate it) affects the resulting extra SSE.
Suppose, for example, we choose $X_{2\backslash 01}=X_{2|01} = (\id - \P_{(X_0,X_1)})X_2$.  Then $X_0$, $X_{1|0}$, and $X_{2|01}$ are pairwise orthogonal matrices and
\[ \P_X - \P_{(X_0, X_{2|01})} = \P_{X_{1|0}},
\]
and the resulting numerator SS is precisely the Type II SS.

Another possibility is to define $X_{2\backslash 01}$ to comprise a set of columns of $X_2$ whose span completes $\sp(X_0, X_1)$ to $\sp(X)$ as a direct sum.  Each possible choice renders $X_{1|0}\bm{\beta}_1$ estimable in its parameterization of $\sp(X)$, and different choices yield different numerator SSs of the test statistic.  Different choices result in different null spaces in $\sp(X)$; while they may all look like they are testing the same hypothesis, they are actually testing different hypotheses in terms of the mean vector: effects of $X_1$ adjusted for $X_0$ are implicitly defined differently. This may be clearer when it is noted that, in each version of the model, $X_{1|0}\bm{\beta}_1$ is the (generally non-orthogonal) projection of $\bm{\mu} = X\bm{\beta}$ onto $\sp(X_{1|0})$ along $\sp(X_0, X_{2\backslash 01})$.

\section{Illustration: Two-Factor ANOVA Model}\label{Illustration}

The Type III construction is illustrated here in the  all-cells-filled (unless specifically noted otherwise), two-factor ANOVA model.  Responses $y_{ijs}$ from subjects $s=1, \ldots, n_{ij}$ ($n_{ij} > 0$) are taken at each of the $ab$ combinations of levels $i= 1, \ldots,a$ of factor A and $j=1, \ldots, b$  of factor B. 

Denote the $ab$-vector of population cell means by $\bm{\eta}=(\eta_{ij})$.  Let $n_{i\cdot} = \sum_j n_{ij}$ and $n_{\cdot\cdot} = \sum_{i,j} n_{ij}$.  Factor A \emph{marginal means} are $\bar{\eta}_{i\cdot} = (1/b)\sum_j\eta_{ij}$, $i=1, \ldots, a$; B marginal means are $\bar{\eta}_{\cdot j} = (1/a)\sum_i\eta_{ij}$, $j=1, \ldots, b$; and $\bar{\eta}_{\cdot\cdot} = (1/ab)\sum_{ij}\eta_{ij}$. 

Let $\mbk$ denote an $n_{\cdot\cdot} \times ab$ matrix. In the row corresponding to the $ijs$-th observation, it has a 1 in the $i,j$-th column and zeroes in all the other columns.  Thus each row of $\mbk$ has exactly one 1, and the $i,j$-th column has $n_{ij}$ 1s. Denote the $n_{\cdot\cdot}$-vector of population means of the response by $\bm{\mu}$.  Then $\bm{\mu}=\mbk\bm{\eta}$.  The $ab$ columns of $\mbk$ are linearly independent (because all $n_{ij}>0$), and so $\sp(\mbk \pr) = \Re^{ab}$: that is, all linear functions of $\bm{\eta}$ are estimable.

The dummy-variable formulation of the model for the cell means follows from the representation
\[ \eta_{ij} = \eta_0 + \alpha_i + \beta_j + \gamma_{ij}.
\]
With the cell means listed in lexicographic order, this becomes
\begin{eqnarray} \bm{\eta} &=& (\eta_{ij}, i=1, \ldots, a, j=1, \ldots, b)\nonumber\\ &=& (\eta_0 + \alpha_i + \beta_j + \gamma_{ij})\nonumber\\ &=& (\bm{1}_a\otimes \bm{1}_b) \eta_0 + (\id_a\otimes \bm{1}_b)\bm{\alpha} + (\bm{1}_a\otimes \id_b)\bm{\beta} + (\id_a\otimes \id_b)\bm{\gamma},
\end{eqnarray}
where $\bm{\alpha} = (\alpha_1, \ldots, \alpha_a)\pr$, $\bm{\beta}=(\beta_1, \ldots, \beta_b)\pr$, and $\bm{\gamma} = (\gamma_{11}, \ldots, \gamma_{ab})\pr$. The model for the mean vector then takes the form
\begin{eqnarray}
\bm{\mu}& =& \mbk \bm{\eta} = \mbk(\bm{1}_a\otimes \bm{1}_b, \id_a\otimes \bm{1}_b, \bm{1}_a\otimes \id_b, \id_a\otimes \id_b)\bm{\theta}, \label{dv_model}
\end{eqnarray}
where $\bm{\theta}$ denotes the vector concatenating $\eta_0$, $\bm{\alpha}$, $\bm{\beta}$, and $\bm{\gamma}$. That is, $\bm{\mu}$ is a vector in $\sp(X)$ with $X=\mbk(\bm{1}_a\otimes \bm{1}_b, \id_a\otimes \bm{1}_b, \bm{1}_a\otimes \id_b, \id_a\otimes \id_b)$.

Denote the Kronecker product matrices in parentheses in (\ref{dv_model}) by $E_{00}$, $E_{10}$, $E_{01}$, and $E_{11}$. They correspond to the intercept term (00), A (10)and B (01) main effects, and AB interaction effects (11).

Define orthogonal projection matrices corresponding to classical ANOVA definitions of effects, as follows.  For each positive integer $m$, define $U_m = (1/m)\bm{1}_m\bm{1}_m\pr$ and $S_m = \id_m - U_m$. For an $m$-vector $\bm{z}$, $U_m\bm{z}=\bar{z}\bm{1}_m$, where $\bar{z} = \bm{1}_m\pr\bm{z}/m$, and $S_m\bm{z} = (z_i - \bar{z})$. 
Let $H_{00} = U_a\otimes U_b$, $H_{10} = S_a\otimes U_b$, $H_{01} = U_a\otimes S_b$, and $H_{11} = S_a\otimes S_b$.  Note that $H_{00}\bm{\eta} = (\bar{\eta}_{\cdot\cdot}, i=1, \ldots, a, j=1, \ldots, b)$, $H_{10}\bm{\eta} = (\bar{\eta}_{i\cdot}-\bar{\eta}_{\cdot\cdot})$, $H_{01}\bm{\eta} = (\bar{\eta}_{\cdot j}-\bar{\eta}_{\cdot\cdot})$, and $H_{11}\bm{\eta} = (\eta_{ij} - \bar{\eta}_{i\cdot} - \bar{\eta}_{\cdot j} + \bar{\eta}_{\cdot\cdot})$.  Note further that the four $H$ matrices are symmetric, idempotent, and pairwise orthogonal, and that their sum is $\id_{ab}$.  

Note that linear subspaces spanned by column-wise concatenations of sets of $E$ matrices are spanned by sums of $H$ matrices. For example, $\sp(E_{00}, E_{10}) = \sp(H_{00}+H_{10})$, and $\sp(E_{00},E_{10}, E_{01}) = \sp(H_{00} + H_{10} + H_{01}) = \sp(\id_{ab}-H_{11})$.

Main effects of factor A are defined as differences among the A level marginal means $\bar{\eta}_{i\cdot}$. Thus there are no A main effects iff $H_{10}\bm{\eta} = \bm{0}$.  

The full model for $\bm{\mu}$ is $\sp(X) = \sp(\mbk)$.  The restricted model for testing A main effects is $\{\mbk\bm{\eta}: \bm{\eta}\in\Re^{ab} \text{ and } H_{10}\bm{\eta}= \bm{0}\}$.  
Because $\{\bm{\eta}: H_{10}\bm{\eta}=\bm{0}\} = \sp(\id_{ab} - H_{10})$, it follows that the restricted model for A main effects is $\sp[\mbk(\id_{ab}-H_{10})]$.  The RMFM numerator SS for A main effects is then
\[  SS_A = \bm{y}\pr[\P_{\mbk} - \P_{\mbk(\id - H_{10})}]\bm{y} = \bm{y}\pr P_{A} \bm{y}.
\]
It is straightforward to show that the ncp $(\mbk\bm{\eta})\pr P_A(\mbk\bm{\eta})/\sigma^2$ is 0 iff $H_{10}\bm{\eta} = \bm{0}$: that is, $SS_A$ tests exactly H$_0: H_{10}\bm{\eta} = \bm{0}$. 

Turning to the Type III construction for A main effects, containment relations are defined in terms of the strings of letters that form the names of the effects. They are that (1) is contained in A, B, and AB, A is contained in AB, and B is contained in AB. 

These define the Type III partition of $X$ into $(X_0, X_1, X_2)$.  With A effects as the target, the effects in the model that do not contain A effects are (1) and B.  The effects that contain A effects are AB.  Then $X_0 = \mbk(\bm{1}_a\otimes \bm{1}_b, \bm{1}_a\otimes \id_b)$, $X_1 = \mbk(\id_a\otimes \bm{1}_b)$, and $X_2 = \mbk(\id_a\otimes \id_b)$.  First find a matrix $N_{01}$ such that $\sp(N_{01}) = \sp(X_0, X_1)^\perp \cap \sp(\mbk)$.  Let $D_{ab} = \Diag(1/n_{ij}) = (\mbk\pr \mbk)^{-1}$. Note that $\sp(X_0, X_1) = \sp[\mbk(\id-H_{11})]$ (the additive-effects model), and that $\sp[\mbk(\id-H_{11})]^\perp \cap\sp(\mbk) = \sp(\mbk D_{ab} H_{11})$.  Then we may choose $N_{01} = \mbk D_{ab} H_{11}$.  Then $X_{2*} = X_2 X_2\pr N_{01} = \mbk H_{11}$;  and $\sp(X_0, X_{2*}) = \sp[\mbk(\bm{1}_a\otimes \bm{1}_b, \bm{1}_a\otimes \id_b, H_{11})] = \sp[\mbk(H_{00} + H_{01} + H_{11})] = \sp[\mbk(\id-H_{10})]$.  

The Type III SS for A main effects is then $\bm{y}\pr P_{A3} \bm{y}$, where $P_{A3}$ is the orthogonal projection matrix onto $\sp(X_0, X_{2*})^\perp \cap\sp(\mbk)$, which is 
\begin{equation}\label{P_3}
P_{A3} = \P_{\mbk} - \P_{(X_0, X_{2*})} = \P_{\mbk} - \P_{\mbk(\id-H_{10})} = P_A.
\end{equation}
From this it is clear that the Type III SS for A main effects is the same as the RMFM SS, and that therefore it tests exactly H$_0: H_{10} \bm{\eta} = \bm{0}$, that the A marginal means are equal. 

The following proposition is useful in establishing the relation between the Type III SS and Yates's MWSM SS. Proof is left to the reader.

\begin{propn}
Let $R$ be an $r\times c$ matrix; $M$, a matrix such that $\sp(M)=\sp(R)^\perp$; $D$, an $r\times r$ symmetric positive-definite (pd) matrix; $D^{1/2}$, a symmetric pd matrix such that $D^{1/2}D^{1/2}=D$; and $D^{-1/2}=(D^{1/2})^{-1}$. Then
\[  \P_{D^{1/2}R} = \id - \P_{D^{-1/2}M}.
\]\label{prop1}
\end{propn}

Next show that $P_A = \P_{\mbk D_{ab}(S_a\otimes \bm{1}_b)}$:
\begin{eqnarray}
\P_{\mbk} - \P_{\mbk(\id-H_{10})} &=& \mbk D_{ab}\mbk\pr - \mbk(\id-H_{10})[(\id-H_{10})D_{ab}^{-1}(\id-H_{10})]^-(\id-H_{10})\mbk\pr\nonumber\\
&=& \mbk D_{ab}^{1/2}(\id - \P_{D_{ab}^{-1/2}(\id-H_{10})})D_{ab}^{1/2}\mbk\pr\nonumber\\
&=& \mbk D_{ab}^{1/2}\P_{D_{ab}^{1/2}H_{10}}D_{ab}^{1/2}\mbk\pr \text{ by Prop. \ref{prop1}}\nonumber\\
&=& \P_{\mbk D_{ab}H_{10}} = \P_{\mbk D_{ab}(S_a\otimes \bm{1}_b)}\label{HtoS}
\end{eqnarray}
 because $\sp(H_{01}) = \sp(S_a\otimes \bm{1}_b)$.

The MWSM SS for A main effects can be expressed as
\begin{equation}\label{MWSM}
Q_{AY} = \bm{u}\pr(D_a^{-1} - D_a^{-1}\bm{1}_a(\bm{1}_a\pr D_a^{-1}
\bm{1}_a)^{-1})\bm{1}_a\pr D_a^{-1})\bm{u},
\end{equation}
(the subscript $Y$ is intended to signify Yates), where $(1/b)\bm{u}$ is the $a$-vector of ``marginal means of the subclass means'' (Yates 1934), where 
\[ \bm{u}=(b\bar{\bar{y}}_i) = \left(\sum_j\bar{y}_{ij\cdot}\right) = (\id_a\otimes \bm{1}_b\pr)\bar{\bm{y}},
\]
and $\bar{\bm{y}} = D_{ab}\mbk\pr \bm{y}$ is the $ab$-vector of cell sample means, which Yates called ``the subclass means.''
$D_a$ is the variance-covariance matrix of $\bm{u}$; it is
\[ D_a=(\id_a\otimes \bm{1}_b)\pr D_{ab}(\id_a\otimes \bm{1}_b) = \Diag\left(\sum_j(1/n_{ij})\right).
\]
In terms of $\bm{y}$, $Q_{AY} = \bm{y}\pr P_{AY}\bm{y}$, where $P_{AY}$ can be re-expressed as:
\begin{eqnarray}
P_{AY} &=& \mbk D_{ab}(\id_a\otimes \bm{1}_b)[D_a^{-1} - D_a^{-1}\bm{1}_a(\bm{1}_a\pr D_a^{-1}\bm{1}_a)^{-1}\bm{1}_a\pr D_a^{-1}](\id_a\otimes \bm{1}_b\pr)D_{ab}\mbk\pr .
\end{eqnarray}

With this notation and formulation it is possible to establish that $P_{A3}$ ($= P_A$) is the same as the $P_{AY}$.  From (\ref{MWSM}),
\begin{eqnarray}
D_a^{-1} - D_a^{-1}\bm{1}_a(\bm{1}_a\pr D_a^{-1}
\bm{1}_a)^{-1})\bm{1}_a\pr D_a^{-1} &=& D_a^{-1/2}(\id-\P_{D_a^{-1/2}\bm{1}_a})D_a^{-1/2}\nonumber\\
&=& D_a^{-1/2}\P_{D^{1/2}S_a}D_a^{-1/2} \text{ by Prop. \ref{prop1}}\nonumber \\
&=& S_a(S_aD_a S_a)^-S_a,\nonumber
\end{eqnarray}
and thus
\begin{eqnarray}
P_{AY} &=& \mbk D_{ab}(\id_a\otimes \bm{1}_b) S_a(S_a D_a S_a)^-S_a(\id_a\otimes \bm{1}_b\pr)D_{ab}\mbk\pr \nonumber\\
&=&  \P_{\mbk D_{ab}(S_a\otimes \bm{1}_b)} .
\end{eqnarray}
The last step follows because $(\id_a\otimes \bm{1}_b)S_a = S_a\otimes \bm{1}_b$, and
\begin{eqnarray*}
[\mbk D_{ab}(S_a\otimes \bm{1}_b)]\pr [\mbk D_{ab}(S_a\otimes \bm{1}_b)] &=& (S_a\otimes \bm{1}_b\pr)D_{ab}(S_a\otimes \bm{1}_b)\\
&=& S_a(\id_a\otimes \bm{1}_b\pr)D_{ab}(\id_a\otimes \bm{1}_b)S_a\\
&=& S_aD_aS_a.
\end{eqnarray*}
Then that $P_3 = P_{AY}$ follows from (\ref{HtoS}).  This establishes that in the two-factor setting with the saturated model for the cell means ($\bm{\eta}\in\Re^{ab}$) and no empty cells, the three SSs for A main effects (RMFM, Type III, and MWSM) are the same.

Searle (1971, p. 371) showed that the MWSM SS for A tests exactly equality of the A marginal means. 
Searle, Speed, and Henderson  (1981, Appendix B)  proved, ``after some tedious algebra,''  that the MWSM SS is equivalent to a SS computed from a set of A contrasts $W\pr \bar{\bm{y}}$ on the cell sample means $\bar{\bm{y}}$ as $SS=(W\pr \bar{\bm{y}})\pr (W\pr D_{ab} W)^- (W\pr\bar{\bm{y}})$. This seems to have been the first algebraic demonstration of the connection between the MWSM SS and any other forms of test statistics. 

If there are any empty cells, the expression for $P_A$ is the same, but it has fewer than $a-1 = \tr(H_{10})$ degrees of freedom.  The MWSM SS is not defined.  The Type III SS is, and it has degrees of freedom equal to the Type II degrees of freedom, as noted in the previous section.  For a single empty cell, it has $a-1$ degrees of freedom, while the RMFM SS has $a-2$, and so the Type III SS and the RMFM SS are not the same.

\section{Bibliography}
\begin{itemize}\setlength{\topsep}{0cm}\setlength{\labelsep}{0cm}
\setlength{\itemsep}{0cm}\setlength{\parsep}{0cm}
\setlength{\parskip}{0cm}\setlength{\itemindent}{-1cm}
\item[] Fisher, R. A. (1938).  Statistical Methods for Research Workers, 7th Edition.  Oliver and Boyd, London.
\item[] Goodnight, J. H. (1976). The General Linear Models procedure. Proceedings of the First International SAS User's Group.  SAS Institute Inc., Cary, NC.
\item[] Hector, A., von Felten, S., Schmid, B. (2010).  Analysis of variance with unbalanced data: an update for ecology \& evolution.  Journal of Animal Ecology 79: 308-316.
\item[] Kutner, M. H. (1974).  Hypothesis testing in linear models (Eisenhart Model I).  The American Statistician, 28(3): 98-100.
\item[] LaMotte, L. R. (2014). The Gram-Schmidt construction as a basis for linear models.  The American Statistician 68: 52-55.
\item[] Langsrud, \O. (2003).  ANOVA for unbalanced data: Use Type II instead of Type III sums of squares.  Statistics and Computing 13:163-167.
\item[] Macnaughton, D. B. (1998).  Which sums of squares are best in unbalanced analysis of variance? MatStat Research Consulting Inc.
\item[] Milliken, G. A., Johnson, D. E. (1984). Analysis of Messy Data, Volume 1: Designed Experiments. Van Nostrand Reinhold Company, New York.
\item[] SAS Institute Inc. (1978).  SAS Technical Report R-101, Tests of hypotheses in fixed-effects linear models.  Cary, NC.
\item[] Searle, S. R., Speed, F. M., and Henderson, H. V. (1981). Some computational and model equivalences in analyses of variance of unequal-subclass-numbers data.  The American Statistician 35: 16-33.
\item[] Smith, C. E., Cribbie, R.  (2014).  Factorial ANOVA with unbalanced data: A fresh look at the types of sums of squares.  Journal of Data Science 12: 385-404.
\item[] Venables, W. N. (2000).  Exegeses on linear models.  Paper presented to the S-Plus User's Conference, Washington, DC, 8-9th October, 1998. \\
\verb#https://www.stats.ox.ac.uk/pub/MASS3/Exegeses.pdf#
\item[] Yates, F. (1934).  The analysis of multiple classificatioins with unequal numbers in the different classes.  Journal of the American Statistical Association, 29(185): 51-66.

\end{itemize}

\end{document}